\documentclass[12pt,prd,preprint,nofootinbib]{revtex4-1}

\usepackage[pdftex]{graphicx}
\usepackage{latexsym,amsmath,amssymb,lmodern,float,url}
\usepackage{natbib}
\usepackage{slashed}
\usepackage{braket}
\usepackage{subfig}
\usepackage[pdftex,bookmarks,linktocpage,pdfpagelabels,plainpages=false,hyperfigures,linkcolor=blue,citecolor=blue]{hyperref} 
\hypersetup{colorlinks=true}

\newcommand\ee{\end{equation}}
\newcommand\be{\begin{equation}}
\newcommand\eea{\end{eqnarray}}
\newcommand\bea{\begin{eqnarray}}

\begin{document}
\bibliographystyle{abbrv}

\title{Extracting Particle Physics Information from Direct Detection of Dark Matter with Minimal Assumptions }
\author{Lawrence M.~Krauss$^{\bf a,b}$}
\author{Jayden L.~Newstead$^{\bf a}$}

\affiliation{$^{\bf a}$ Department of Physics, Arizona State University, Tempe, AZ 85287, USA,}
\affiliation{$^{\bf b}$ School of Earth and Space Exploration, Arizona State University, Tempe, AZ 85287, USA,}

\bigskip

\begin{abstract}

In the absence of direct accelerator data to constrain particle models, and given existing astrophysical uncertainties associated with the phase space distribution of WIMP dark matter in our galactic halo, extracting information on fundamental particle microphysics from possible signals in underground direct detectors will be challenging. Given these challenges we explore the requirements for direct detection of dark matter experiments to extract information on fundamental particle physics interactions.   In particular, using Bayesian methods, we explore the quantitative distinctions that allow differentiation between different non-relativistic effective operators, as a function of the number of detected events, for a variety of possible operators that might generate the detected distribution. Without a spinless target one cannot distinguish between spin-dependent and spin-independent interactions. In general, of order 50 events would be required to definitively determine that the fundamental dark matter scattering amplitude is momentum independent, even in the optimistic case of minimal detector backgrounds and no inelastic scattering contributions.  This bound can be improved with reduced uncertainties in the dark matter velocity distribution.   
\end{abstract}

\maketitle

\section{Introduction}

While the existence of dark matter is well established~\cite{Bertone:2004pz}, its composition and origin are not. Perhaps the simplest, and most well-motivated, particle physics candidates are weakly interacting massive particles (WIMPs) which were thermally produced in the early universe. Directly detecting WIMPs in the galactic dark matter halo through elastic nuclear scattering is thus the goal of many present experimental efforts~\cite{Undagoitia:2015gya}. 

While it is known that direct detection experiments will have some ability to distinguish different WIMP scattering models~\cite{Peter:2013aha,Catena:2014uqa,Dent:2015zpa,Gluscevic:2015sqa,Baum:2017kfa,Kahlhoefer:2016eds} a detailed quantitative analysis of the requirements for this to be achievable under realistic circumstances and with minimal theoretical assumptions has not been fully explored. In this work we extend previous analyses by making as few assumptions as possible about the WIMP model and astrophysics while placing our emphasis on model selection, restricting our analysis to a subset of relevant operators.  Assuming the anticipated sensitivity of the next generation of direct detectors, we find that in order to obtain statistically unambiguous conclusions allowing distinction between different particle physics models, given our current limited knowledge of the astrophysical parameters of dark matter, the minimum number of detected events ranges from a low of 40 events, up to several hundred events, depending upon the model.

Our paper is organized as follows, the WIMP models to be considered are discussed in Sec.~\ref{secModels}, the WIMP velocity distribution and how we reconstruct it is presented in Sec.~\ref{secVel}. The details of how we carry out Bayesian model selection and the results are given in~\ref{secMS}, and finally in Sec.~\ref{secDC} we discuss the results and provide some concluding remarks.\\

\section{WIMP models}
\label{secModels}
\begin{figure}
\centering
\subfloat[][]{\includegraphics[width=7cm]{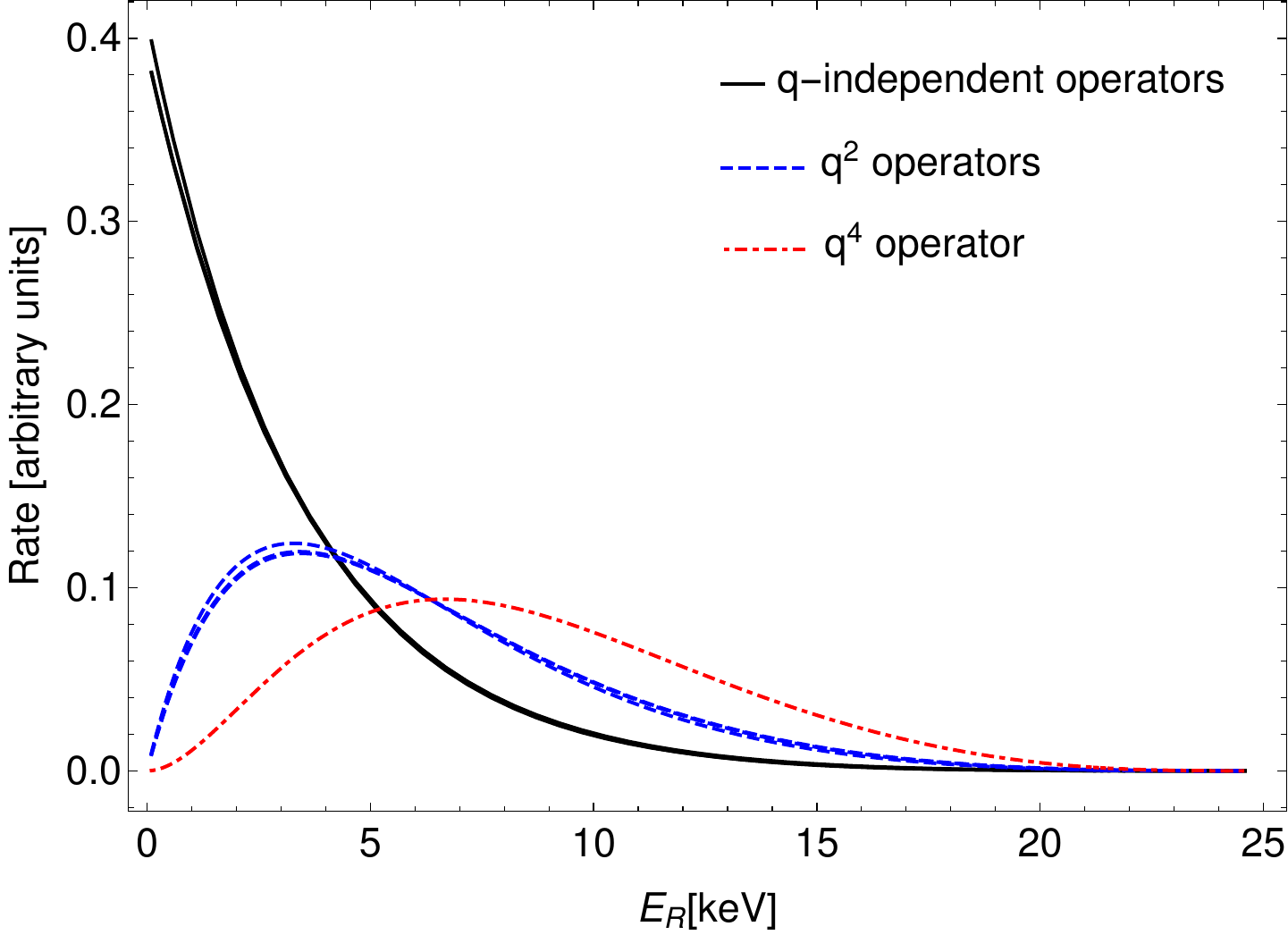}}\quad
\subfloat[][]{\includegraphics[width=7cm]{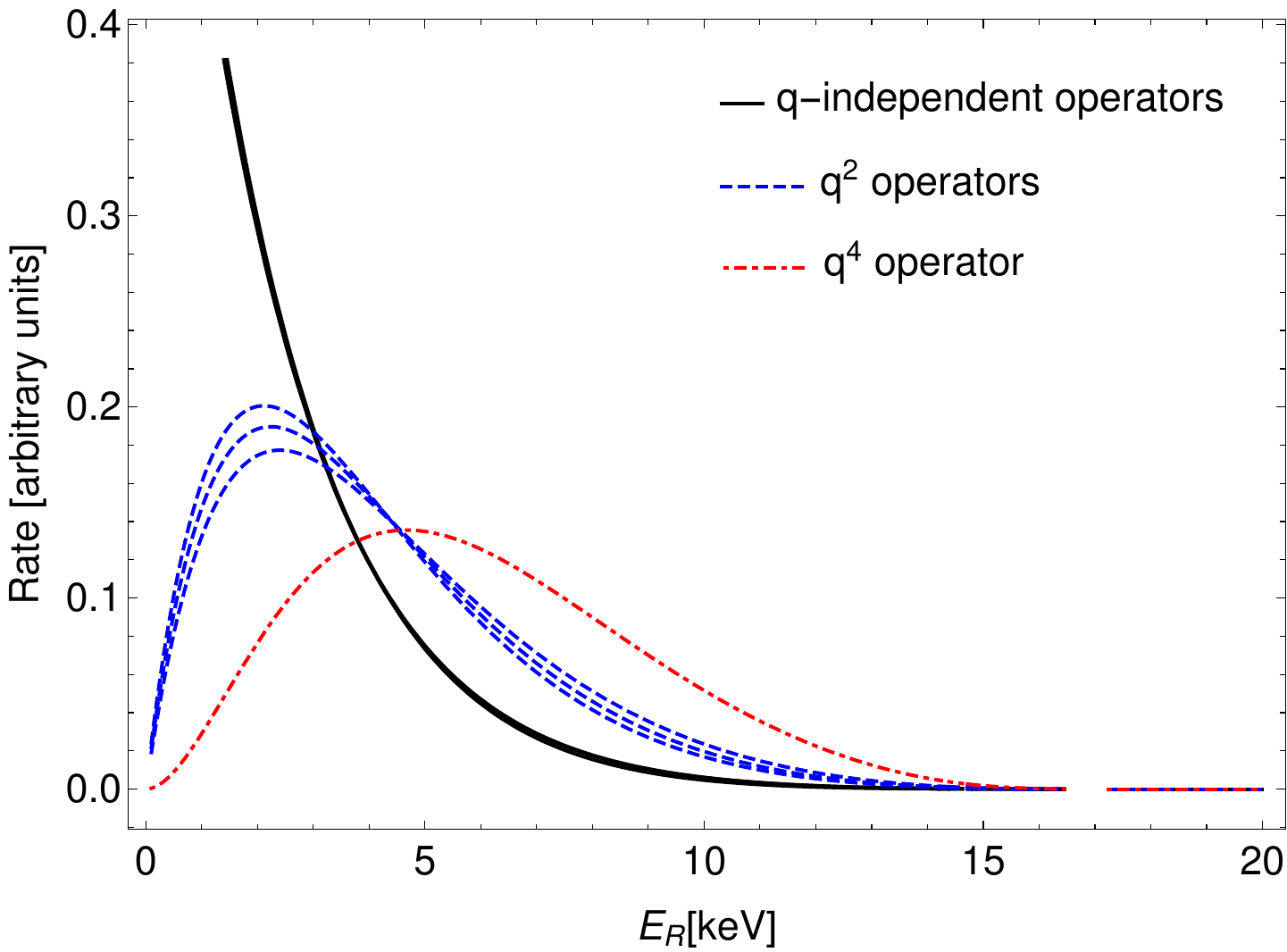}}
\caption{Sample rates for a 20 GeV WIMP in germanium (left) and xenon (right) for each of the operators in Eq.~\ref{eqnOperators}, normalized to produce the same total number of events in a given exposure between $0.1-30$ keV (germanium) and $1-20$ keV (xenon)}
\label{figRates}
\end{figure}

Given that WIMPs are assumed to be heavy decoupled particles, it is natural to describe their interactions using non-relativistic effective field theories~\cite{Fan:2010gt,Fitzpatrick:2012ix,Anand:2014kea}. Fitzpatrick et al. constructed a complete list from all the relevant variables in the problem; the momentum transfer, $q$, the velocity $v^\perp$, the WIMP spin $\vec{S}_\chi$ and the nuclear spin, $\vec{S}_N$. Scalar combinations of these vectors (up to second order in $q$) gives a set of 15 independent operators, plus two additional operators if considering vector WIMPs~\cite{Dent:2015zpa}. However, when considering fermionic WIMPs in a simple UV model\footnote{in a 'simple model' we require the WIMP to couple to quarks via a single mediator with dimension-4 interaction terms}, only 7 operators can be produced at leading order~\cite{Dent:2015zpa}: 
\bea
\mathcal{O}_1 &=& 1_\chi1_N, \nonumber\\
\mathcal{O}_4 &=& \vec{S}_\chi\cdot\vec{S}_N\nonumber\\
\mathcal{O}_6 &=& (\frac{\vec{q}}{m_N}\cdot\vec{S}_\chi)(\frac{\vec{q}}{m_N}\cdot\vec{S}_N) ,\nonumber\\
\mathcal{O}_8 &=& \vec{v}^\perp\cdot\vec{S}_\chi,\nonumber\\
\mathcal{O}_9 &=& i\vec{S}_\chi\cdot(\vec{S}_N\times\frac{\vec{q}}{m_N}),\nonumber\\
\mathcal{O}_{10} &=& i\frac{\vec{q}}{m_N}\cdot\vec{S}_N,\nonumber\\
\mathcal{O}_{11} &=& \frac{\vec{q}}{m_N}\cdot\vec{S}_\chi\nonumber\\
\label{eqnOperators}
\eea
 In \cite{Fitzpatrick:2012ix} these operators have been decomposed into the standard electroweak response functions, and nuclear form factors based on these respones are then provided for various nuclei. Large scale nuclear structure calculations are notoriously difficult and the results can be highly dependent on the method used~\cite{Vietze:2014vsa}. Ab initio calculations are possible, but not yet practical for large nuclei~\cite{Gazda:2016mrp}. These potentially large systematic uncertainties can further confuse our ability to distinguish WIMP candidates. Therefore to limit the impact of nuclear effects on the WIMP scattering rates we will consider low momentum transfers. In this region differential event rates are depend primarily on momentum dependence rather than the details of nuclear structure.  Operators can then be classified based based on their resulting momentum dependence, either $1,q,$ or $q^2$. In the low momentum transfer region, the differential rates for the operators in each group become essentially degenerate, as demonstrated in Fig.~\ref{figRates}. For the detectors considered here, this region corresponds to WIMP masses below around 25 GeV.  We can also allow for inelastic scattering, which can produce additional momentum dependent effects. It should be stressed that if the nuclear structure was well known, the high-momentum transfer recoils would provide an extra factor with which to distinguish between operators. To avoid fitting with incorrect nuclear structure, we will consider the challenge of distinguishing the momentum dependence of the scattering operator for light WIMPs. 

In this analysis we label each model, $M_i$, in terms of WIMP scattering via a single NR operator $\mathcal{O}_i$. The parameters of each model are the WIMP mass, operator coefficient, inelastic parameter and isospin violating parameter: $\theta_i = {m_\chi,c_i,\delta,f_n/f_p}$.
 To simplify our analysis, we consider here 4 representative operators, one momentum independent, one $q$-dependent and one $q^2$-dependent ($\mathcal{O}_1$, $\mathcal{O}_{10}$ and $\mathcal{O}_{6}$ respectively). In Sec.\ref{secMS} we will determine how many events are required to distinguish these operators. Note, that operators within a given momentum group are not able to be distinguished based on recoil rate alone. To explore this degeneracy we also include the standard spin-dependent operator $\mathcal{O}_{4}$, which has the same momentum dependence as $\mathcal{O}_1$. The scattering rate in a detector is found by integrating over the incoming WIMP velocities,
\bea
\frac{dR}{dE_R} = N_T \frac{\rho_\chi M}{2\pi m_\chi}\int_{v_{min}}\frac{f(v)}{v} P_{tot} dv
\eea
where $\rho_\chi=0.3$GeV/cm$^3$ is the local dark matter density, $N_T$ is the number of target nuclei and $P_{tot}$ is calculated from the amplitude $\mathcal{M}$,
\bea
P_{tot} = \frac{1}{2j_\chi+1}\frac{1}{2j_N+1}\sum_{spins}|\mathcal{M}|^2.
\eea
For the details of the computation of the amplitude see~\cite{Fitzpatrick:2012ix}, throughout this work we use the Mathematica package supplied in~\cite{Anand:2013yka} to calculate rates.

\section{Velocity distribution}
\label{secVel}
Evaluation of the rates requires integration over the WIMP velocity distribution. Empirically there is no direct probe of the velocity distribution, and thus it's form is not well constrained. This has led to the development of generalized astrophysical independent methods, which are useful for evaluating the compatibility of experimental results~\cite{Fox:2010bz,Gondolo:2012rs}. They are not well suited to parameter estimation problems since, by design, they do not directly constrain WIMP parameters. In this work we will use the generalized distribution formalism due to Green and Kavanaugh~\cite{Kavanagh:2013wba}. It has been shown that this method can reconstruct various WIMP velocity distributions without biasing the derived WIMP mass~\cite{Kavanagh:2013eya}. In this formalism the velocity distribution is parametrized by the exponent of a polynomial series,
\be
f(v) = v^2 \mathrm{Exp}\left[-\sum_{i=0} a_i \mathrm{P}_i(\frac{v}{v_{\mathrm{max}}})\right]
\ee
where the $P_i(x)$ are any set of orthogonal polynomials.  Here we will use shifted Chebyshev polynomials, which have been shown to have desirable properties~\cite{Kavanagh:2013eya}. We truncate the sum at 5 Chebyshev polynomials, giving us a balance between flexibility in the reconstruction and computational time.
In order to produce a simulated experimental dataset for use in the reconstruction analysis we choose a velocity distribution, which for simplicity choose to be a Maxwell-Boltzman distribution cutoff at the escape velocity, which, in the galactic frame, is given by,
\be
f(\vec{v}) = \frac{1}{n}\left[e^{\frac{v^2}{v_0^2}}-e^{\frac{v_{\mathrm{esc}}}{v_0^2}}\right]\Theta(v_{\mathrm{esc}}^2-\vec{v}^2).
\ee
Here $v_0=220$ km/s is the local circular velocity, $v_{\mathrm{esc}}=544$ km/s is the local escape velocity, and $n$ is a normalization factor. Finally, the distribution must be boosted into the laboratory frame by the Earth's velocity, $\vec{v}_e=\vec{v}_0+\vec{v}_{\odot\mathrm{pec}}+\vec{v}_\oplus$.

\section{Model selection}
\label{secMS}
Bayesian statistical methods are a natural choice for parameter estimation problems, but they also simultaneously provide a method for model selection through the use of Bayesian Evidence. The Evidence for a model $M_i$, given some experimental data $D$, will be expressed as $\epsilon(D,M_i)$, and can be found from Bayes theorem,
\be
\mathcal{P}(\theta_i,M_i|D) = \frac{\mathcal{L}(D|\theta_i,M_i)\pi(\theta_i,M_i)}{\epsilon(D,M_i)}.
\ee
Here the likelihood function, $\mathcal{L}$, is taken to be a binned poisson likelihood, and the prior probabilities $\pi(\theta_i,M_i)$ are given in table \ref{tabPar}. Since the posterior probability distribution is normalized to unity, it is possible to determine the Evidence via an integral over the model parameter space, $\theta_i$,
\be
\epsilon(D,M_i) = \int \mathcal{L}(D|\theta_i,M_i)\pi(\theta_i,M_i) d\theta_i.
\ee
 When performing model selection with Bayesian statistics the figure of merit considered is the Bayes factor, $K$, which is the ratio of model evidences:  
\be
K = \frac{\epsilon(D,M_i)}{\epsilon(D,M_j)}.
\ee
Also known as an odds ratio, the Bayes factor allows us to state how much more likely $M_i$ is able to reproduce the data, in comparison with $M_j$. This is in contrast to the Bayes factor used in \cite{Gluscevic:2015sqa}, which compares a model $M_1$ with all competing hypotheses. Jeffreys considered a $K>100$ as decisive evidence for $M_i$ over $M_j$~\cite{Jeffreys61a}, while others have used the criterion $2\mathrm{ln}(K)>10$ \cite{kass:1995}. We use this more stringent requirement of $\mathrm{ln}(K)>5$ when concluding that an experiment could definitively discern the difference beween two models. \\

\begin{table}[bt]
\centering
\caption{Parameter space and priors for the reconstruction, the velocity priors are only used in the Maxwell-Boltzmann reconstructions}
\begin{tabular}{|c|c|c|}
\hline
Parameter & Range             & Prior \\
\hline
$m_\chi$  & $1-10^3$ GeV       & log \\
$c_1$     &  $10^{-10}-10^4$ GeV$^{-2}$  & log \\
$c_4$     &  $10^{-10}-10^4$ GeV$^{-2}$ & log \\
$c_6$     &  $10^{-10}-10^4$ GeV$^{-2}$ & log \\
$c_{10}$  &  $10^{-10}-10^4$ GeV$^{-2}$ & log \\
$f^n/f^p$ & $-4 - 4$                & linear \\
$\delta$ &  $0-100$ keV             & linear \\
$\rho_\chi$ & $0.3\pm0.1$GeV/cm$^3$ & gaussian \\
$a_i$ & $-50 - 80$ & linear \\
\hline
\hline
$v_0$ & $220\pm 20$ km/s & gaussian \\
$v_\mathrm{esc}$ & $544\pm 40$ km/s & gaussian \\
\hline
\end{tabular}
\label{tabPar}
\end{table}

Each model was simulated to produce an increasing number of events (from 10 to 250) in both xenon and germanium detectors. To explore the most optimistic possibility for detection, we assumed a WIMP mass of 20 GeV throughout and assumed and neither isospin violation nor inelastic scattering. The coefficients and exposure were arbitrarily scaled to achieve the desired number of events within the energy ranges 0.1-30keV and 1-20keV for germanium and xenon respectively. We choose to work with the expected (Asimov) dataset, to reduce computing requirements. Events were binned into 20 equally spaced intervals in these ranges. Both detectors were assumed to be background free and no detector resolution effects were included.

\begin{figure}[ht!]
\centering
 \subfloat[][]{\includegraphics[width=7cm]{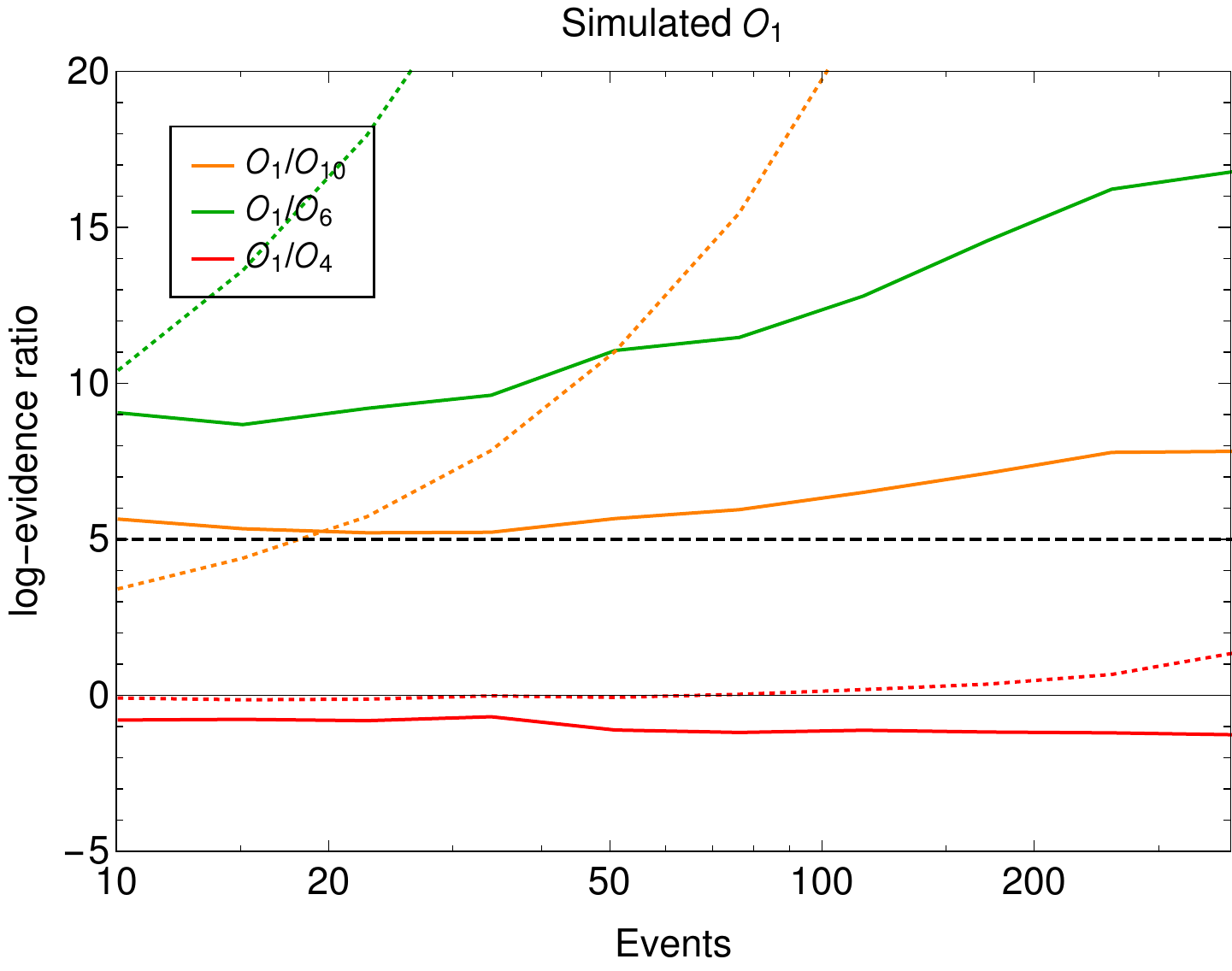}}\quad
 \subfloat[][]{\includegraphics[width=7cm]{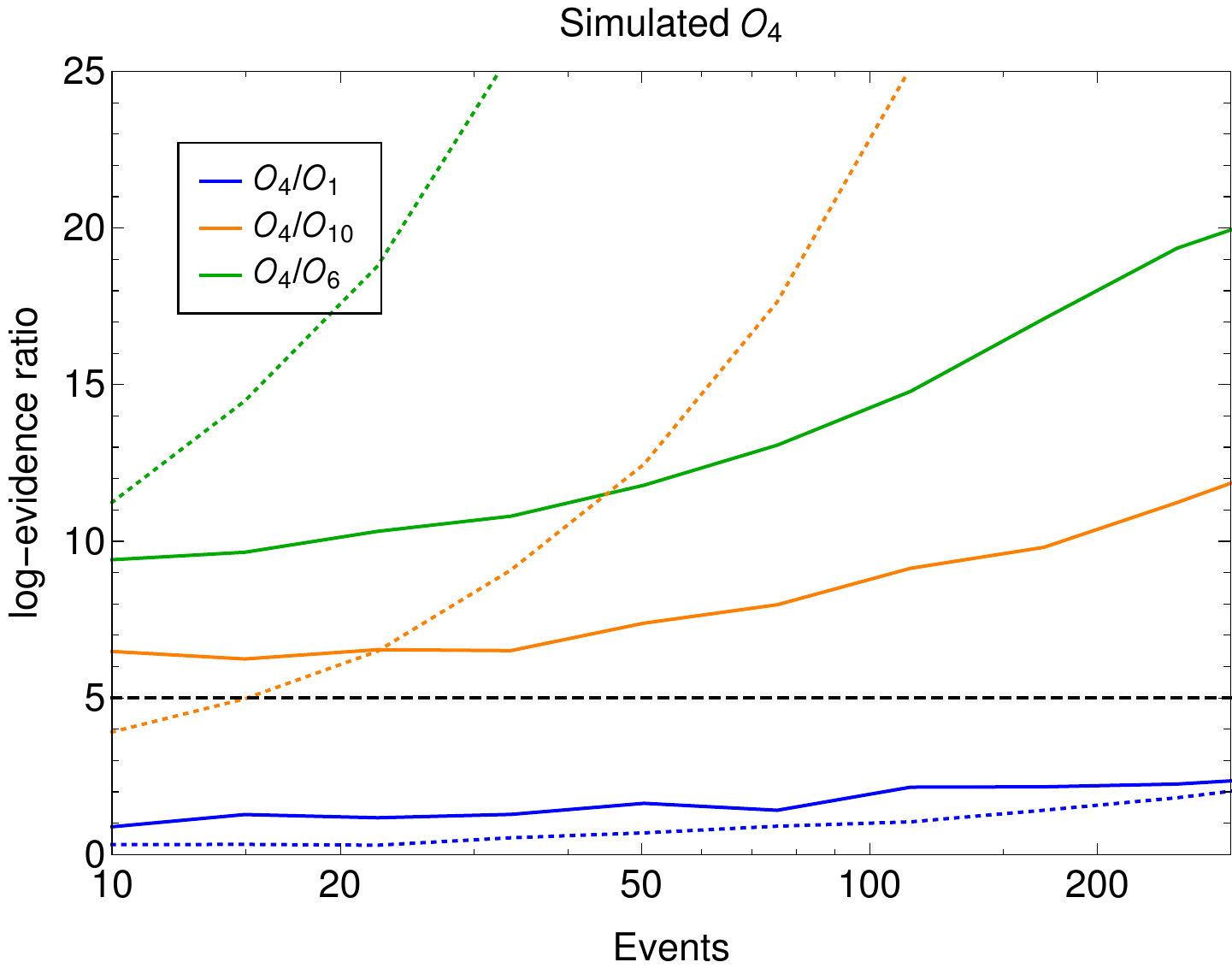}} \\
 \subfloat[][]{\includegraphics[width=7cm]{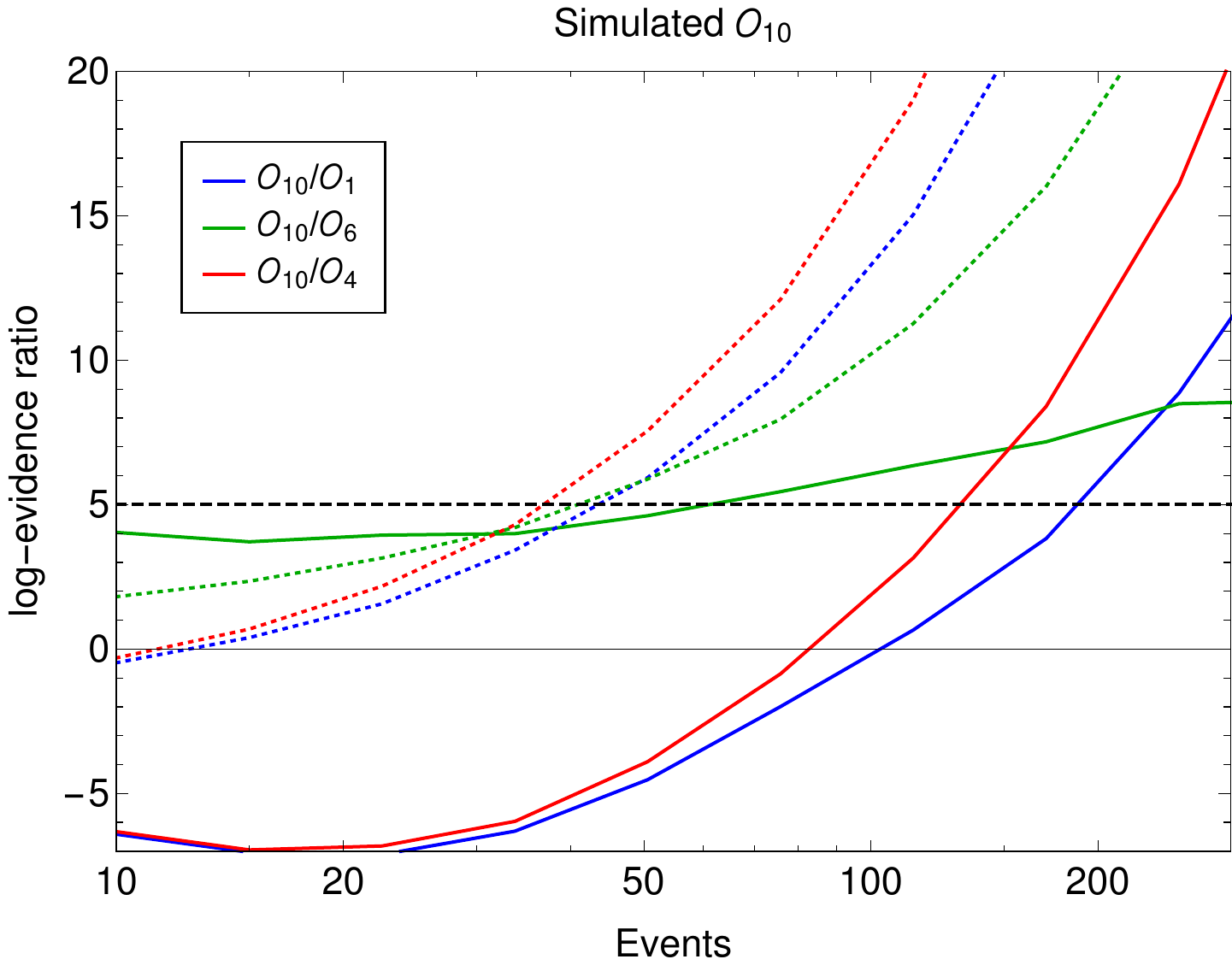}}\quad
 \subfloat[][]{\includegraphics[width=7cm]{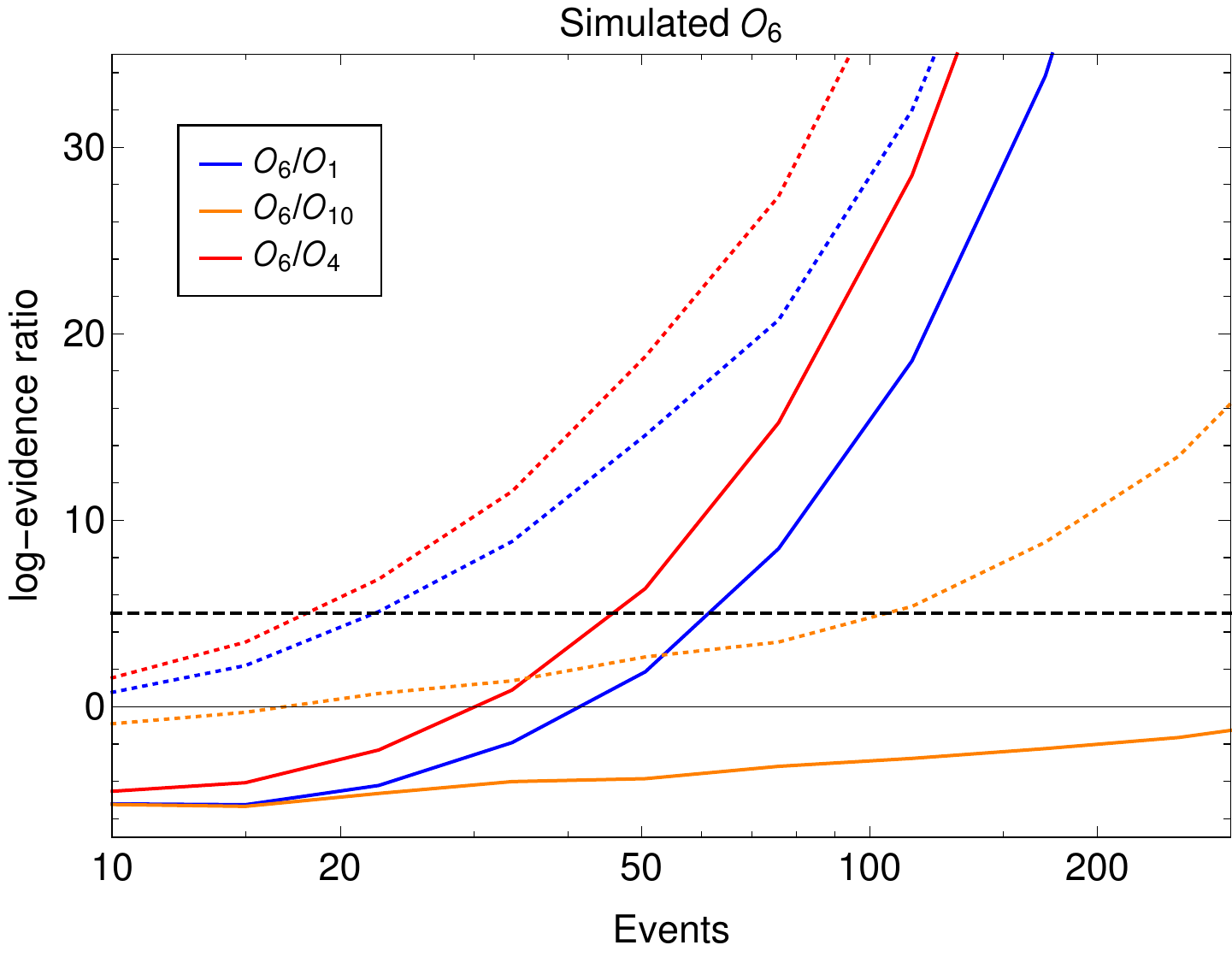}}
\caption{The evidence ratios for the four models under the simulation of $\mathcal{O}_1$ (top left), $\mathcal{O}_4$ (top right), $\mathcal{O}_{10}$ (bottom left) and $\mathcal{O}_6$ (bottom right). The solid (dashed) lines indicate that the Bayes factors were calculated with the generalized (Maxwell-Boltzman) velocity distribution}
\label{figEv}
\end{figure}

The MultiNest code, described in \cite{multinest08}, was used to sample the posterior probability distribution and calculate the model evidence over the priors listed in table~\ref{tabPar}. For each simulation Bayes factors were calculated for each model verses the simulated model. This allowed us to see how the Bayes factors evolve as a function of number of detected events, shown in Fig.~\ref{figEv}. Since we choose to form our Bayes factors with the simulated model in the numerator and we do not have statistical fluctuations, these functions are monotonically increasing as more events are collected. For comparison, the procedure was repeated assuming a fixed Maxwell-Boltzmann velocity distribution with priors on the velocities (given in table~\ref{tabPar}).

\section{Discussion and conclusion}
\label{secDC}
The simplest possible fundamental interaction to probe is model $M_1$ corresponding to the spin independent operator $\mathcal{O}_1$, and implying no momentum-dependence in the rate and if we simulate events based on this model, then only a handful of events will be required to discern that this is the case. However, allowing for the possibility of isospin violation, to distinguish this model from $M_4$ would require use of a third detector beyond xenon and germanium, which are both sensitive to spin-dependent interactions.  A detector not sensititve to spin (e.g. argon) could then trivially break this degeneracy by reduced rate in the case of interactions arising from $M_4$. 

Once we consider interactions with momentum-dependence in the rate, the situation becomes more confused, however.  If $M_{10}$ is the true model we find that as few as 50 events and as many as 200 events are required to distinguish between possible operators as the source of detected events. Curiously, in the case of very few events, the ability to vary the generalized velocity distribution can cancel out inferred momentum-dependence in the rate, so that for few events, $M_1$ and $M_4$ can be favored over the correct model ($M_{10}$). Thus, in the case of few events, a large Bayes factor is not a guarantee that the correct underlying dark matter model has been identified.

When $M_{6}$ is the true model we find that the q-independent models ($M_1$ and $M_4$) are ruled out after around 50 events, but without velocity priors, degeneracy with $M_{10}$ remains beyond 250 events. 
In almost all cases the addition of velocity priors allows one to distinguish between models with fewer events. However in some cases, when there are a low number of events, larger Bayes factors are again a red herring, as the additional freedom allowed by the generalized velocity distribution without priors can enhance the apparent agreement with data, even if the data is generated using the restricted Maxwell-Boltzmann distribution.  

In short, in the case of few events, any model inferences based on statistical fits to data are suspect, at least until astrophysical uncertainties in the dark matter velocity distribution are reduced.  This is particularly important if one is to distinguish model interactions that result in momentum-independent rates from momentum-dependent rates. 

While positive detection of particle dark matter in our galactic halo would be a discovery of profound importance for physics, our analysis demonstrates the challenges associated with distinguishing fundamental particle interaction models for dark matter that might be possible on the basis of observed signals in direct detection experiments,--even under very optimistic assumptions about experimental noise, and possible complications like inelastic scattering--will be considerable.  On the order of 30-50 events will be necessary before any reliable distinction between fundamental models will be possible.   In order to make this estimate firmer, in future work we will incorporate additional possible confusion factors, from inelastic scattering to detector noise.     We will also explore the quantitative effect of reducing dark matter velocity distribution uncertainties on the ability to distinguish particle models. 

We thank the DOE and ASU for financial support for this research.

\bibliography{phys}

\end{document}